\documentclass[12pt,a4paper,twosided]{article}
\newenvironment{changemargin}[2]{%
\begin{list}{}{%
\setlength{\leftmargin}{#1}%
\setlength{\rightmargin}{#2}%
}%
\item[]}
{\end{list}}
\usepackage{graphicx}
\usepackage{color}
\usepackage[noadjust]{cite}
\usepackage{lscape}
\usepackage{epstopdf}
\usepackage{lscape}
\usepackage{amsmath}
\usepackage{amsmath}
\usepackage{setspace}

\usepackage{fullpage}
\usepackage{geometry}
\usepackage{hyperref}
\usepackage{booktabs} 
\usepackage{array} 
\usepackage{paralist}
\usepackage{verbatim} 
\usepackage{subfigure} 
\usepackage{bm}
\begin{document}
\baselineskip=0.27in

{\bf \LARGE
\begin{changemargin}{-0.5cm}{-0.5cm}
\begin{center}
{Quantum teleportation and information splitting via four-qubit cluster state and a Bell state}
\end{center}\end{changemargin}}
\vspace{4mm}
\begin{center}
{\large{\bf Marlon David Gonz\'alez Ram\'irez$^{a,}$}}\large{\bf ,} \large{\bf Babatunde James Falaye$^{b,c,}$}\footnote{\scriptsize E-mail:~ fbjames11@physicist.net}\large{\bf ,} {\large{\bf Guo-Hua Sun$^{d,}$}}\large{\bf ,} {\large{\bf M. Cruz-Irisson$^{a,}$}} \large{\bf and} {\large{\bf Shi-Hai Dong$^{e,}$}}
\end{center}
{\footnotesize
\begin{center}
{\it $^\textbf{a}$ESIME-Culhuacan, Instituto Polit\'ecnico Nacional, Av. Santa Ana 1000, Mexico, CDMX 04430, Mexico}\\
{\it $^\textbf{b}$Departamento de F\'isica, Escuela Superior de F\'isica y Matem\'aticas, Instituto Polit\'ecnico Nacional, Edificio 9, UPALM, CDMX 07738, M{e}xico}\\
{\it $^\textbf{c}$Applied Theoretical Physics Division, Department of Physics, Federal University Lafia,  P. M. B. 146, Lafia, Nigeria}\\
{\it $^\textbf{d}$C{a}tedr\'{a}tica CONACyT, CIC, Instituto Polit\'{e}cnico Nacional, UPALM, CDMX 07700, M{e}xico}\\
{\it $^\textbf{e}$Laboratorio de Informaci\'on Cu\'antica, CIDETEC, Instituto Polit\'{e}cnico Nacional, UPALM, CDMX 07700, M{e}xico.}
\end{center}}
\begin{abstract}
\noindent
Quantum teleportation provides a `bodiless' way of transmitting the quantum state from one object to another, at a distant location, using a classical communication channel and a previously shared entangled state. In this paper, we present a tripartite scheme for probabilistic teleportation of an arbitrary single qubit state, without losing the information of the state being teleported, via a four-qubit cluster state of the form $\left.|{\phi}\right\rangle_{1234}=\left.\alpha|0000\right\rangle+\left.\beta|1010\right\rangle+\left.\gamma|0101\right\rangle-\eta\left.|1111\right\rangle$, as the quantum channel, where the nonzero real numbers $\alpha$, $\beta$, $\gamma$, and $\eta$ satisfy the relation $|\alpha|^2+|\beta|^2+|\gamma|^2+|\eta|^2=1$. With the introduction of an auxiliary qubit with state $\left|0\right\rangle$, using a suitable unitary transformation and a positive-operator valued measure (POVM), the receiver can recreate the state of the original qubit. An important advantage of the teleportation scheme demonstrated here is that, if the teleportation fails, it can be repeated without teleporting copies of the unknown quantum state, if the concerned parties share another pair of entangled qubit. We also present a protocol for quantum information splitting of an arbitrary two-particle system via the aforementioned cluster state and a Bell-state as the quantum channel. Problems related to security attacks were examined for both the cases and it was found that this protocol is secure. This protocol is highly efficient and easy to implement.
\noindent
\end{abstract}

{\bf Keywords}: Teleportation; Positive-operator valued measure; Tripartite scheme; 

Security attacks.

{\bf PACs No.}: 03.65.Ud, 03.67.Hk, 03.67.Ac, 03.67.Mn.

\section{Introduction}
\label{sec1}
Suppose when you visit your friend’s home you find a very fine chair, and you wish to acquire a similar one for your sitting room. One of the strategic plans that will come to your mind is to measure all the properties of this chair and then reproduce a copy for your sitting room. However, if we intend to use the same strategy for a quantum particle, e.g. photon, it will fail, because of the no-cloning theorem of quantum mechanics. Then, what could be done in this situation? The answer is simply quantum teleportation. This idea, which was put forward by Bennett et al. \cite{A1}, involves the transfer of the unknown properties of an object, say $A$ onto second object, say $B$ without disobeying the no-cloning theorem, which states that ``it is impossible to perfectly copy properties of quantum particles." The misconception about teleportation is that the transfer happens instantaneously. However, it follows processes. Firstly, during quantum teleportation, the object $A$ losses all its properties and then they are transferred. After this transfer, the properties of object $B$ remain unknown. All that could be known is that object $B$ is now identical to object $A$ before teleportation.

Thus, quantum teleportation can be described as a process by which quantum information can be transmitted from one location to another, with the help of classical communication and previously shared quantum entanglement between the sending and receiving location. Quantum teleportation requires that a classical information link must be set, in order to allow the transmission of classical bits to accompany each qubit. After the first teleportation protocol was proposed by Bennett et al., many researchers have proposed several quantum teleportation protocols (\cite{A2,A3,A4,R1,R2,R3,A5,A6,A7,A8} and Refs. therein). Some of these protocols have been realized experimentally (\cite{TY1} and Refs. therein). The two key deterministic factors in teleporting the qubit of an unknown state are the two classical bits and maximally entangled Einstein-Podolsky-Rosen pairs, which act as the quantum channel. The application of quantum entanglement is not limited to quantum teleportation; it is also being employed in quantum information splitting, quantum dense coding, quantum secret sharing, quantum cloning, etc.

It is worth mentioning that quantum entanglements have been realized experimentally. Some of the entangled states include: Greenberger-Horne-Zeilinger (GHZ), $W$ class, and the cluster state. GHZ is fully separable after the loss of one qubit, unlike W-class, which will still be entangled with the remaining two qubits. The GHZ of three photons and three Rydberg atoms have been observed experimentally \cite{B2}, while an entangled $W$ state of three qubits have been realized in Ref \cite{B3}. In addition, the cluster state \cite{B4}, which is a highly entangled state of multiple qubits, is generated in lattices of qubits with Ising type interactions. On the basis of single qubit operation, the cluster state serves as the initial resource of a universal computation scheme \cite{B5}. The cluster state has been realized experimentally in photonic experiments \cite{B5} and in optical lattices of cold atoms \cite{B6}. 

Motivated by the unending interest in studying quantum teleportation, in this paper, we demonstrate a scheme for probabilistic teleportation of an arbitrary single qubit state, using a four-qubit cluster state as the quantum channel (Section 2). Furthermore, we also demonstrate quantum information splitting of an arbitrary two-particle system in Section 3. To achieve these objectives, we assume that Alice, Bob, and Chika are the concerned parties in a quantum network. It should be noted that the names we have adopted here are for convenience. For instance, saying “Alice sends a message to Chika through Bob” is easier to follow than saying “party A sends a message to party C through party B.” The concluding remarks are given in Section 4.
 
\section{Probabilistic teleportation of an arbitrary single qubit state}
We shall demonstrate the usefulness of the cluster state $\left.|{\phi}\right\rangle_{1234}=\left.\alpha|0000\right\rangle+\left.\beta|1010\right\rangle+\left.\gamma|0101\right\rangle-\eta\left.|1111\right\rangle$ as the quantum channel for the teleportation of a single qubit system. For $\alpha=\beta=\gamma=\eta=1/2$, this entangled qubit is among the three ultimate maximally entangled states identified by Gour and Wallach \cite{B1}. It has a property such that for 2 out of the $3$ bipartite cuts, the entanglement is $2$ ebits, and for the last bipartite cut, the entanglement between the groups of two qubits is $1$ ebit. The nonzero real numbers $\alpha$, $\beta$, $\gamma$ and $\eta$ satisfy the relation $|\alpha|^2+|\beta|^2+|\gamma|^2+|\eta|^2=1$.

Now, suppose that an arbitrary single qubit state that Alice wishes to teleport to Chika, who is in a distant location, under the control of Bob is $\left.|\check{\phi}\right\rangle_{\mathcal{A}}=\left.a_0|0\right\rangle_{\mathcal{A}}+\left.b_0|1\right\rangle_{\mathcal{A}}$, where $a_0$ and $b_0$ are complex and satisfy the condition $|a_0|^2+|b_0|^2=1$. Assuming that Alice, Bob, and Chika shared a four-qubit cluster state, where qubit 1 belongs to Alice, qubits $2$ and $3$ belong to Bob, and the fourth to Chika.The state of the total system is expressed as $\left.|\check{\Phi}\right\rangle_{\mathcal{A}1234}=\left.|\check{\phi}\right\rangle_{\mathcal{A}}\otimes\left.|{\phi}\right\rangle_{1234}$.

In order to achieve the quantum teleportation, Alice performs Bell state measurement (BSM) on her qubit pair $({\mathcal{A}},1)$, and consequently, the combined state of the other qubits collapse to four possible pure entangled three-qubit states after the measurement. Afterwards, Alice utilizes the classical channel to publish her results. Now, suppose Bob agrees to assist Chika in reconstructing the original state of the qubit. He then performs a BSM on his qubit pair $(2, 3)$, and consequently, the state of fourth qubit becomes:
\begin{subequations}
\begin{eqnarray}
&& _{2,3}\left\langle\Phi^{\pm}\right.|\ _{\mathcal{A},1}\left\langle\Phi^{\pm}\right.|\left.\check{\Phi}\right\rangle_{\mathcal{A}1234}=\frac{1}{2}\left[\left.a_0\alpha|0\right\rangle_{4}\pm^{(\mathcal{A},1)}\mp^{(2,3)}\left.b_0\eta|1\right\rangle_{4}\right]\label{EQ2a}, \\
&& _{2,3}\left\langle\Psi^{\pm}\right.|\ _{\mathcal{A},1}\left\langle\Phi^{\pm}\right.|\left.\check{\Phi}\right\rangle_{\mathcal{A}1234}=\frac{1}{2}\left[\left.\pm^{(2,3)}b_0\beta|0\right\rangle_{4}\pm^{(\mathcal{A},1)}\left.a_0\gamma|1\right\rangle_{4}\right]\label{EQ2b},\\
&& _{2,3}\left\langle\Phi^{\pm}\right.|\ _{\mathcal{A},1}\left\langle\Psi^{\pm}\right.|\left.\check{\Phi}\right\rangle_{\mathcal{A}1234}=\frac{1}{2}\left[\left. \pm^{(2,3)}b_0\alpha|0\right\rangle_{4}\mp^{(\mathcal{A},1)}\left.a_0\eta|1\right\rangle_{4}\right]\label{EQ2c}, \\
&& _{2,3}\left\langle\Psi^{\pm}\right.|\ _{\mathcal{A},1}\left\langle\Psi^{\pm}\right.|\left.\check{\Phi}\right\rangle_{\mathcal{A}1234}=\frac{1}{2}\left[\left.a_0\beta|0\right\rangle_{4}\pm^{(\mathcal{A},1)}\pm^{(2,3)}\left.b_0\gamma|1\right\rangle_{4}\right]\label{EQ2d}.
\end{eqnarray}
\end{subequations}
Since equation (\ref{EQ2a})-(\ref{EQ2d}) are not normalized, it then implies that each outcome of the measurement has a different probability. Now, Bob tells Chika the results of his measurement using a classical channel. In order to avoid redundancy, we shall not discuss all the outcomes, but only one. For the other cases, Chika can apply a similar approach to reconstruct the original state. Now, suppose the results of Alice and Bob results are $|\left.\Phi^{+}\right\rangle_{\mathcal{A},1}|\left.\Psi^{-}\right\rangle_{2,3}$. Consequently, without loss of generality, the state of qubit $4$ collapses to $\mathcal{Q}=2^{-1}(\left.a_0\gamma|1\right\rangle_{4}-\left.b_0\beta|0\right\rangle_{4})$ according to Equation (\ref{EQ2b}). With this result, Chika can recover the original state $\left.|\check{\phi}\right\rangle$.

For Chika to perform this task, first, she needs to set up a close similarity such that the coefficient of $a_0$ should be $\left|0\right\rangle_{4}$ and that of $b_0$ should be $\left|1\right\rangle_{4}$. To achieve this, she performs a local unitary operation $\mathcal{U}_1=\left|0\right\rangle_4\left\langle 1\right|-\left|1\right\rangle_4\left\langle 0\right|$ on $\mathcal{Q}$ and then she introduces an auxiliary qubit, say $\mathcal{T}$ with state $\left.|0\right\rangle_{\mathcal{T}}$, which gives $\mathcal{Q}_1=2^{-1}(\left.a_0\gamma|00\right\rangle_{4\mathcal{T}}+\left.b_0\beta|10\right\rangle_{4\mathcal{T}})$. She then performs a C-NOT operation on qubit pair ($4,\mathcal{T}$) to obtain
\begin{equation}
\mathcal{Q}'_1=\frac{1}{4}\left[\left(\left.a_0|0\right\rangle_{4}+\left.b_0|1\right\rangle_{4}\right)\otimes\left(\left.\gamma|0\right\rangle_{\mathcal{T}}+\left.\beta|1\right\rangle_{\mathcal{T}}\right)+\left(\left.a_0|0\right\rangle_{4}-\left.b_0|1\right\rangle_{4}\right)\otimes\left(\left.\gamma|0\right\rangle_{\mathcal{T}}-\left.\beta|1\right\rangle_{\mathcal{T}}\right)\right].
\label{EQ3}
\end{equation}
As evident from Equation (\ref{EQ3}), with the condition that $\left.\gamma|0\right\rangle_{\mathcal{T}}+\left.\beta|1\right\rangle_{\mathcal{T}}$ and $\left.\gamma|0\right\rangle_{\mathcal{T}}-\left.\beta|1\right\rangle_{\mathcal{T}}$ can be discerned using a suitable measurement,  Chika can obtain $\left.a_0|0\right\rangle_{4}+\left.b_0|1\right\rangle_{4}$ or $\left.a_0|0\right\rangle_{4}-\left.b_0|1\right\rangle_{4}$. In order to achieve this discrimination, Chika needs to perform POVM \cite{A9,A10} on the auxiliary qubit $\mathcal{T}$.

POVM is a measure with non-negative self-adjoint operators on a Hilbert space, and its integral is represented by the identity operator. A common mistake is to confuse POVM with measurement outcomes. The measurement outcomes are arbitrary and rely on a scale (for instance, position relies on a frame of reference); therefore, one can fix that scale differently and change the measurement outcomes. The necessity for the POVM arises due to the projective measurements on a larger system, through which the measurements that are performed mathematically by a projection-valued measure will act on a subsystem in such a way that cannot be described by a projection-valued measure on the subsystem alone. Thus, this POVM, which Chika needs to adopt, can be written mathematically as:
\begin{equation}
\mathcal{K}_1=\frac{1}{{\varrho}}\left.|M_1\right\rangle\left\langle M_1|\right.,\ \ \ \mathcal{K}_2=\frac{1}{{\varrho}}\left.|M_2\right\rangle\left\langle M_2|\right.,\ \ \ \mathcal{K}_3=I-\frac{1}{\varrho}\sum_{i=1}^2\left.|M_i\right\rangle\left\langle M_i|\right.\label{EQ4},
\end{equation}
where
\begin{equation}
\left.|M_1\right\rangle=\frac{1}{\sqrt{\varsigma}}\left(\frac{1}{\gamma}\left.|0\right\rangle+\frac{1}{\beta}\left.|1\right\rangle\right),\ \ \ \left.|M_2\right\rangle=\frac{1}{\sqrt{\varsigma}}\left(\frac{1}{\gamma}\left.|0\right\rangle-\frac{1}{\beta}\left.|1\right\rangle\right),\ \ \mbox{and}\ \ \varsigma=\frac{1}{\gamma^2}+\frac{1}{\beta^2}.\label{EQ5}
\end{equation}
$I$ denotes an identity operator and $\varrho$ is a parameter, which defines the range of positivity of the operator $\mathcal{K}_3$. Thus, $\varrho$ must lie within 1 and 2 (i.e. $1\leq\varrho\leq2$) for $\mathcal{K}_3$ to be a nonnegative operator. Now, suppose Chika's POVM result on qubit $\mathcal{T}$ yields $\mathcal{K}_1$, whose probability is $\left\langle\mathcal{Q}_1\right.|\mathcal{K}_1|\left.\mathcal{Q}_1\right\rangle=1/(4\varrho\varsigma)$, then she can deduce the state of qubit $4$ to be $\left.a_0|0\right\rangle_{4}+\left.b_0|1\right\rangle$, which is $\left.|\check{\phi}\right\rangle_{\mathcal{A}}$, i.e., the original state. In this case, Chika need not do anything. However, suppose the result is $\mathcal{K}_2$, with a probability calculated by $\left\langle\mathcal{Q}_1\right.|\mathcal{K}_2|\left.\mathcal{Q}_1\right\rangle=1/(4\varrho\varsigma)$, then she finds that the state of qubit $4$ is $\left.a_0|0\right\rangle_4-\left.b_0|1\right\rangle_4$. In this case, Chika can fix up her state by applying an appropriate unitary transformation. 

In the three-dimensional visualization of a qubit, the spin-up state $\left.|0\right\rangle_{4}$ denotes the north pole, while the spin-down state $\left.|1\right\rangle_4$ denotes the south pole. Thus, a superposition represents any state in between the poles. A phase shift with $\theta=\pi$ leaves the basis state $\left.|0\right\rangle_4$ unchanged and maps $\left.|1\right\rangle_4$ to -$\left.|1\right\rangle_4$. This phase shift is denoted by the Pauli-$z$ matrix ($\sigma^z$). With the application of this operation on the state of qubit $4$, i.e., $\sigma^z(\left.a_0|0\right\rangle_4-\left.b_0|1\right\rangle_4)$, yields $\left.a_0|0\right\rangle+\left.b_0|1\right\rangle=\left.|\check{\phi}\right\rangle$, which was the original state that was teleported.

Suppose the result is $\mathcal{K}_3$, with a probability given as $ 1-1/(2\varrho\varsigma)$, then the scheme fails due to the fact that it is impossible for Chika to deduce the state of qubit $4$. Thus, the probability for success is $1/(2\varrho\varsigma)$. However, suppose Alice, Bob, and Chika share another qubit pair. Then, the process can be repeated when the teleportation fails, provided Alice, Bob, and Chika share another pair of the entangled qubits. Thus, we obtain the sum of the probabilities that there are $M$ successes when $N$ independent teleportation trials are carried out as
\begin{equation}
P_{suc}=\sum_{M=1}^{N-1}\left(\begin{matrix}N-1\\M\end{matrix}\right)\frac{(2\varrho\varsigma-1)^{N-M-1}}{(2\varrho\varsigma)^{N}}.\label{EQ6}
\end{equation}
\begin{figure*}[!t]
\centering \includegraphics[height=100mm, width=140mm]{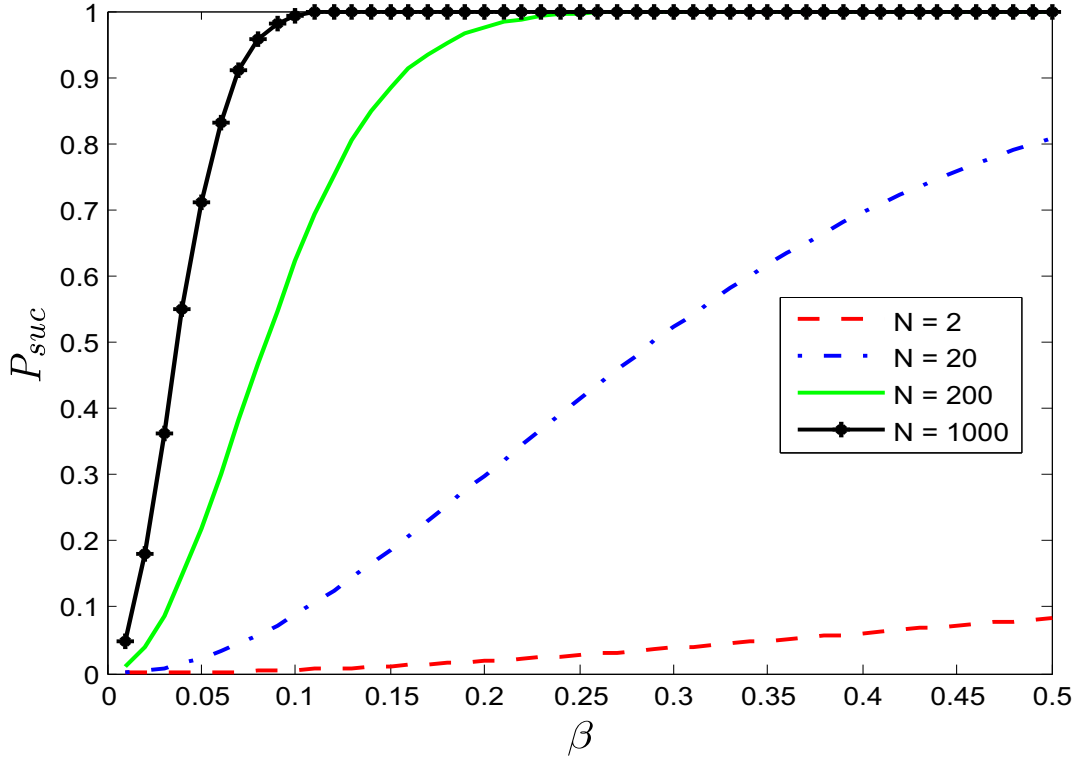}
\caption{\protect\footnotesize Variation of the probability given by equation (\ref{EQ6}) as a function of $\beta$. We take $\gamma=0.5$ and $\varrho=1.5$. It is shown that for a fixed $\beta$, the success probability increases considerably as $n$ increases.} 
\label{fig1}
\end{figure*}

If we consider a special case $N=2$, our result agrees with Ref. \cite{R4}. Figure \ref{fig1} gives the variation of the chance of success as a function of parameter $\beta$. Now, let us assume that there is an eavesdropper (Eve), who intends to obtain information about the unknown qubit state without the consent of Alice, Bob, and Chika, who are legitimate partners. Assume that she successfully entangles an auxiliary qubit $2^{-1/2}(\left.|0\right\rangle+\left.|1\right\rangle)_E$ to the qubit possessed by Bob. Then, after Alice performs BSM, the combined state of Bob, Chika, and Eve collapses into $\left|\Pi^{(1)}\right\rangle_{234E}=\ _{\mathcal{A},1}\left\langle\Phi^{\pm}\right.|\left.\check{\Phi}\right\rangle_{\mathcal{A}1234}\left[\frac{1}{\sqrt{2}}(\left.|0\right\rangle+\left.|1\right\rangle)_E\right]$ and $\left|\Pi^{(2)}\right\rangle_{234E}=\ _{\mathcal{A},1}\left\langle\Phi^{\pm}\right.|\left.\check{\Psi}\right\rangle_{\mathcal{A}1234}\left[\frac{1}{\sqrt{2}}(\left.|0\right\rangle+\left.|1\right\rangle)_E\right].$When Bob performs BSM on his qubits pair $(2,3)$, he obtains 
\begin{eqnarray}
\left|\Pi^{(1)}\right\rangle_{234E}&=&\frac{1}{2\sqrt{2}}\Bigg[\left.|00\right\rangle\left(a_0\alpha\left.|00\right\rangle\mp b_0\eta\left.|10\right\rangle+a_0\alpha\left.|01\right\rangle\mp b_0\eta\left.|11\right\rangle\right)\nonumber\\
&&\ \ \ \ \ \ +\left.|10\right\rangle\left(a_0\alpha\left.|00\right\rangle+a_0\alpha\left.|01\right\rangle-\left(\mp b_0\eta\left.|10\right\rangle\mp b_0\eta\left.|11\right\rangle\right)\right)\nonumber\\
&&\ \ \ \ \ \ +\left.|01\right\rangle\left(\pm b_0\beta\left.|00\right\rangle+a_0\gamma\left.|10\right\rangle\pm b_0\beta\left.|01\right\rangle+a_0\gamma\left.|11\right\rangle\right)\nonumber\\
&&\ \ \ \ \ \ +\left.|11\right\rangle\left(\pm b_0\beta\left.|00\right\rangle-a_0\gamma\left.|10\right\rangle\pm b_0\beta\left.|01\right\rangle-a_0\gamma\left.|11\right\rangle\right)\Bigg].\nonumber\\\label{EQ8}
\end{eqnarray}
Similar calculations can be repeated for $\left|\Pi^{(2)}\right\rangle_{234E}$,and consequently, the state of Chika-Eve system becomes
\begin{subequations}
\begin{eqnarray}
&& _{2,3}\left\langle\Phi^{\pm}\right|\left.\Pi^{(1)}\right\rangle_{234E}=\frac{1}{2}\left[\left.a_0\alpha|0\right\rangle_{4}\mp^{(\mathcal{A},1)}\pm^{(2,3)}\left.b_0\eta|1\right\rangle_{4}\right]\left[\frac{1}{\sqrt{2}}(\left.|0\right\rangle+\left.|1\right\rangle)_E\right],\label{EQ9a} \\
&& _{2,3}\left\langle\Psi^{\pm}\right|\left.\Pi^{(1)}\right\rangle_{234E}=\frac{1}{2}\left[\left.\pm^{(2,3)}a_0\gamma|1\right\rangle_{4}\pm^{(\mathcal{A},1)}\left.b_0\beta|1\right\rangle_{4}\right]\left[\frac{1}{\sqrt{2}}(\left.|0\right\rangle+\left.|1\right\rangle)_E\right],\label{EQ9b}\\
&& _{2,3}\left\langle\Phi^{\pm}\right|\left.\Pi^{(2)}\right\rangle_{234E}=\frac{1}{2}\left[\left. \mp^{(2,3)}a_0\eta|1\right\rangle_{4}\pm^{(\mathcal{A},1)}\left.b_0\alpha|0\right\rangle_{4}\right]\left[\frac{1}{\sqrt{2}}(\left.|0\right\rangle+\left.|1\right\rangle)_E\right],\label{EE9c} \\
&& _{2,3}\left\langle\Psi^{\pm}\right|\left.\Pi^{(2)}\right\rangle_{234E}=\frac{1}{2}\left[\left.a_0\beta|0\right\rangle_{4}\pm^{(\mathcal{A},1)}\pm^{(2,3)}\left.b_0\gamma|1\right\rangle_{4}\right]\left[\frac{1}{\sqrt{2}}(\left.|0\right\rangle+\left.|1\right\rangle)_E\right].\label{EQ9d}
\end{eqnarray}
\end{subequations}
From Equations (\ref{EQ9a}-\ref{EQ9d}), it is clear that Eve's state is not altered, and consequently, it is impossible for her to obtain any reasonable information about the state of the teleported qubit. Therefore, this protocol is secure.

\section{Quantum information splitting of an arbitrary two-particle state}
Quantum information splitting (QIS) is a technique of sharing and splitting of quantum information among two or more parties such that none of them can retrieve the information fully by operating on his/her own qubits. The QIS involving GHZ as the quantum channel has been extensively studied (\cite{A11} and Refs. therein). Experimentally, QIS has been realized through a pseudo GHZ state \cite{A12}. In addition, experimental demonstration of the four-party quantum secret sharing via the resource of four-photon entanglement has been presented in Ref. \cite{A13}.

In this section, we demonstrate a scheme for splitting an arbitrary two-particle state via four-qubit cluster and Bell state as the quantum channel linking Alice, Bob, and Chika. Now, let us suppose that the splitter (Alice) has an arbitrary two-particle state $\left|{\Psi}^0\right\rangle_{ab}=a_1\left|00\right\rangle+b_1\left|01\right\rangle+c_1\left|10\right\rangle+d_1\left|11\right\rangle$ (where $a_1, b_1, c_1$ and $d_1$ satisfy the condition $|a_1|^2+|b_1|^2+|c_1|^2+|d_1|^2=1$), which she intends to share with Bob and Chika. For Alice to achieve her aim, it is necessary to prepare a four-qubit cluster state and Bell state (i.e., $\left.|{\phi}\right\rangle_{1234}\otimes\left|\Phi^{+}\right\rangle_{56}$) as the quantum channel linking the three parties. Qubits a, b, 1, and 6 belong to Alice while qubits $2$ and $3$ belong to Bob and qubits $4$ and $5$ belong to Chika. Thus, the state of the whole information system can be expressed as $\left|\hat{\Phi}\right\rangle_{ab123456}=\left|{\Psi}^o\right\rangle_{ab}\otimes\left.|{\phi}\right\rangle_{1234}\otimes\left|\Phi^{+}\right\rangle_{56}$.

\begin{figure*}[!t]
\centering \includegraphics[height=80mm, width=120mm]{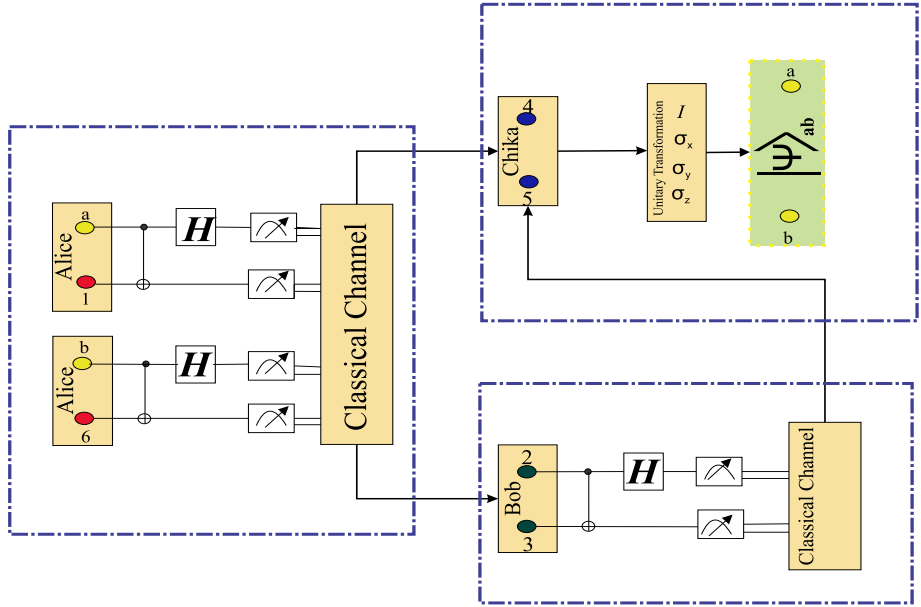}
\caption{\protect\footnotesize Schematic representation of the quantum information splitting of an arbitrary two-particle state via four-qubit cluster and a Bell state as quantum channel linking Alice, Bob, and Chika. The particles $a$, $b$, $1$, and $6$ belong to Alice, while the particles $2$ and $3$ belong to Bob, and the particles $4$ and $5$ belong to Chika. $H$ is the Hadamard gate. $\sigma_x$, $\sigma_y$, and $\sigma_z$ are Pauli matrices. $I$ is the $2\times2$ identity matrix. The meters represent measurement. The double lines coming out of the meters carry classical bits (single lines denote qubits). Alice performs BSM on her qubit pairs $(a,1)$ and $(b,6)$. Alice communicates the results of her measurement to Bob and Chika via the classical channel. Bob performs BSM on his qubits pair $(2,3)$. With the information received from Alice and Bob, Chika can recover the original state of the qubit via an appropriate unitary transformation.}
\label{fig2}
\end{figure*}

Now, for QIS to be achieved, Alice performs BSM on her qubit pairs ($a,1$) and ($b,6$). The combined state of the other qubits (i.e., $16$ possible states, which Alice's system will collapse to after the measurement) with equal probability can be found. Now, Alice communicates her results to Bob and Chika using the classical channel. Neither Bob nor Chika can reconstruct the original state of the qubits if there is no collaboration, since neither of them have enough information to do so. In other words, any of them could reconstruct the original state of the qubits, provided there is collaboration between the two parties. This is the main idea of QIS. Now, let us suppose Bob collaborates with Chika in reconstructing the original state of the qubits. Then, he needs to perform a BSM on his qubits pair $(2, 3)$, and consequently, the states of Chika's qubits become
\begin{subequations}
\begin{equation}
_{2,3}\left\langle\Phi^{\pm}\right|\ _{b,6}\left\langle\Phi^{\pm}\right|\ _{a,1}\left\langle\Phi^{\pm}\right|\left.\hat{\Phi}\right\rangle_{ab123456}=\frac{1}{8}\Bigg[a_1\left|00\right\rangle\pm^{(2)} b_1\left|01\right\rangle\pm^{(1)}\mp^{(3)}c_1\left|10\right\rangle\pm^{(1)}\pm^{(2)}\mp^{(3)} d_1\left|11\right\rangle\Bigg],
\label{EQ11a}
\end{equation}
\begin{equation}
_{2,3}\left\langle\Psi^{\pm}\right|\ _{b,6}\left\langle\Phi^{\pm}\right|\ _{a,1}\left\langle\Phi^{\pm}\right|\left.\hat{\Phi}\right\rangle_{ab123456}=\frac{1}{8}\Bigg[\pm^{(3)}a_1\left|10\right\rangle\pm^{(2)}\pm^{(3)}b_1\left|11\right\rangle\pm^{(1)}c_1\left|00\right\rangle\pm^{(1)}\pm^{(2)}d_1\left|01\right\rangle\Bigg],
\label{EQ11b}
\end{equation}
\begin{equation}
_{2,3}\left\langle\Phi^{\pm}\right|\ _{b,6}\left\langle\Psi^{\pm}\right|\ _{a,1}\left\langle\Phi^{\pm}\right|\left.\hat{\Phi}\right\rangle_{ab123456}=\frac{1}{8}\Bigg[a_1\left|01\right\rangle\pm^{(2)} b_1\left|00\right\rangle\pm^{(1)}\mp^{(3)}c_1\left|11\right\rangle\pm^{(1)}\pm^{(2)}\mp^{(3)}d_1\left|10\right\rangle\Bigg],
\label{EQ11c}
\end{equation}
\begin{equation}
_{2,3}\left\langle\Psi^{\pm}\right|\ _{b,6}\left\langle\Psi^{\pm}\right|\ _{a,1}\left\langle\Phi^{\pm}\right|\left.\hat{\Phi}\right\rangle_{ab123456}=\frac{1}{8}\Bigg[\pm^{(3)}a_1\left|11\right\rangle\pm^{(2)}\pm^{(3)}b_1\left|10\right\rangle\pm^{(1)}c_1\left|01\right\rangle\pm^{(1)}\pm^{(2)}d_1\left|00\right\rangle\Bigg],
\label{EQ11d}
\end{equation}
\begin{equation}
_{2,3}\left\langle\Phi^{\pm}\right|\ _{b,6}\left\langle\Phi^{\pm}\right|\ _{a,1}\left\langle\Psi^{\pm}\right|\left.\hat{\Phi}\right\rangle_{ab123456}=\frac{1}{8}\Bigg[\mp^{(3)}a_1\left|10\right\rangle\pm^{(2)}\mp^{(3)} b_1\left|11\right\rangle\pm^{(1)}c_1\left|00\right\rangle\pm^{(1)}\pm^{(2)} d_1\left|01\right\rangle\Bigg],
\label{EQ11e}
\end{equation}
\begin{equation}
_{2,3}\left\langle\Psi^{\pm}\right|\ _{b,6}\left\langle\Phi^{\pm}\right|\ _{a,1}\left\langle\Psi^{\pm}\right|\left.\hat{\Phi}\right\rangle_{ab123456}=\frac{1}{8}\Bigg[a_1\left|00\right\rangle\pm^{(2)}b_1\left|01\right\rangle\pm^{(1)}\pm^{(3)}c_1\left|10\right\rangle\pm^{(1)}\pm^{(2)}\pm^{(3)}d_1\left|11\right\rangle\Bigg],
\label{EQ11f}
\end{equation}
\begin{equation}
_{2,3}\left\langle\Phi^{\pm}\right|\ _{b,6}\left\langle\Psi^{\pm}\right|\ _{a,1}\left\langle\Psi^{\pm}\right|\left.\hat{\Phi}\right\rangle_{ab123456}=\frac{1}{8}\Bigg[\mp^{(3)}a_1\left|11\right\rangle\pm^{(2)}\mp^{(3)} b_1\left|10\right\rangle\pm^{(1)}c_1\left|01\right\rangle\pm^{(1)}\pm^{(2)} d_1\left|00\right\rangle\Bigg],
\label{EQ11g}
\end{equation}
\begin{equation}
_{2,3}\left\langle\Psi^{\pm}\right|\ _{b,6}\left\langle\Psi^{\pm}\right|\ _{a,1}\left\langle\Psi^{\pm}\right|\left.\hat{\Phi}\right\rangle_{ab123456}=\frac{1}{8}\Bigg[a_1\left|01\right\rangle\pm^{(2)}b_1\left|00\right\rangle\pm^{(1)}\pm^{(3)}c_1\left|11\right\rangle\pm^{(1)}\pm^{(2)}\pm^{(3)}d_1\left|10\right\rangle\Bigg],
\label{EQ11h}
\end{equation}
\end{subequations}
$\pm^{(1)}$, $\pm^{(2)}$ and $\pm^{(3)}(\mp^{(3)})$  represent the results corresponding to the BSM on the qubit pairs ($a,1$), ($b,6$), and ($2,3$) respectively. A more explicit expression for Equations (\ref{EQ11a}-\ref{EQ11h}) and the appropriate unitary transformation, which Chika could utilize in reconstructing the original state, are shown in Table 1. This process is also illustrated in Figure \ref{fig2}.

Let us now make a security analysis of this scheme. For this purpose, we assume that an eavesdropper (Eve) intends to steal information about the state of the unknown qubit state without the consent of Alice, Bob, and Chika, who are legitimate partners. Assuming she successfully entangles an ancilla qubit $2^{-1/2}(\left.|0\right\rangle+\left.|1\right\rangle)_E$ to the qubit possessed by Bob, then after Alice performs BSM on her qubits pair, the combined state of Bob, Chika, and Eve collapses into the five-qubit entangled state. For instance, if Alice obtains $\left|\Phi^{\pm}\right\rangle_{a,1}\left|\Phi^{\pm}\right\rangle_{b,6}$, then the combined state of Bob, Chika, and Eve will be 
\begin{eqnarray}
\left|\Lambda^{(1)}\right\rangle_{2345E}&=&\frac{1}{8}\Bigg[a_1\left(\left.|00000\right\rangle+\left.|10100\right\rangle\right)\pm^{(2)} b_1\left(\left.|00010\right\rangle+\left.|10110\right\rangle\right)\nonumber\\
&&\pm^{(1)} c_1\left(\left.|01000\right\rangle-\left.|11100\right\rangle\right)\pm^{(1)}\pm^{(2)} d_1\left(\left.|01010\right\rangle-\left.|11110\right\rangle\right)\nonumber\\
&&+a_1\left(\left.|00001\right\rangle+\left.|10101\right\rangle\right)\pm^{(2)} b_1\left(\left.|00011\right\rangle+\left.|10111\right\rangle\right)\nonumber\\
&&\pm^{(1)} c_1\left(\left.|01001\right\rangle-\left.|11101\right\rangle\right)\pm^{(1)}\pm^{(2)} d_1\left(\left.|01011\right\rangle-\left.|11111\right\rangle\right)\Bigg]\label{EQ12}. 
\end{eqnarray}
Now, Bob performs BSM on qubits pair $(2,3)$. Suppose Bob obtains $\left|\Phi^{\pm}\right\rangle$, then Chika-Eve system will collapse into 
\begin{eqnarray}
_{2,3}\left\langle\Phi^{+}\right|\left.\Lambda^{(1)}\right\rangle_{2345E}&=&\frac{1}{8\sqrt{2}}\Bigg[a_1\left(\left.|000\right\rangle+\left.|001\right\rangle\right)\pm^{(2)} b_1\left(\left.|010\right\rangle+\left.|011\right\rangle\right)-^{(3)}\pm^{(1)} \nonumber\\
&&c_1\left(\left.|100\right\rangle+\left.|101\right\rangle\right)-^{(3)}\pm^{(1)}\pm^{(2)} d_1\left(\left.|110\right\rangle+\left.|111\right\rangle\right)\Bigg]_{45E}\label{EQ13}\\
&&=\frac{1}{8}\Bigg[a_1\left|00\right\rangle\pm^{(2)} b_1\left|01\right\rangle-^{(3)}\pm^{(1)}c_1\left|10\right\rangle-^{(3)}\pm^{(1)}\pm^{(2)} d_1\left|11\right\rangle\Bigg]_{45}\left[\frac{1}{2}(\left.|0\right\rangle+\left.|1\right\rangle)\right]_E.\nonumber
\end{eqnarray}
It is evident that Eve's state is unaltered, and consequently, it is impossible for her to obtain the original qubit state. Thus, this protocol is secure. Moreover, suppose Bob is dishonest and ready to cooperate with Eve, and in order to intercept the original qubit sent by Alice, he sends the entanglement qubit, which has been prepared beforehand, to Chika. In this situation, Bob will find it difficult to obtain suitable quantum information when she needs to reconstruct the original state of the qubits. Since the information about the quantum state received by Chika from Bob is erroneous, the communications are consequently discarded.
\begin{table}[!t]
{\scriptsize
\caption{\footnotesize Alice's results, the outcome of the measurement performed by Bob, the corresponding state obtained Chika, and the appropriate unitary transformation utilized by Chika to reconstruct the original state of the qubit. We have ignored the normalization for convenience} \vspace*{10pt}{\footnotesize
\begin{tabular}{cccccccccc}\hline\hline
{}&{}&{}&{}&{}&{}&{}&{}&{}&{}\\[-1.0ex]Alice's result&&&Bob's result&&& State obtained by Chika&&& Unitary transformation\\[1ex]\hline
$\left|\Phi^+\right\rangle\left|\Phi^+\right\rangle$&&&$\left|00\right\rangle+\left|11\right\rangle$&&&$
a_1\left|00\right\rangle+b_1\left|01\right\rangle-c_1\left|10\right\rangle-d_1\left|11\right\rangle$&&&$\sigma_z\otimes I$ \\[1ex]
$\left|\Phi^-\right\rangle\left|\Phi^+\right\rangle$&&&$\left|00\right\rangle+\left|11\right\rangle$&&&$	a_1\left|00\right\rangle+b_1\left|01\right\rangle+c_1\left|10\right\rangle+d_1\left|11\right\rangle$&&&$I\otimes I$ \\[1ex]
$\left|\Phi^+\right\rangle\left|\Phi^-\right\rangle$&&&$\left|00\right\rangle+\left|11\right\rangle$&&&$	a_1\left|00\right\rangle-b_1\left|01\right\rangle-c_1\left|10\right\rangle+d_1\left|11\right\rangle$&&&$\sigma_z\otimes\sigma_z$ \\[1ex]
$\left|\Phi^-\right\rangle\left|\Phi^-\right\rangle$&&&$\left|00\right\rangle+\left|11\right\rangle$&&&$	a_1\left|00\right\rangle-b_1\left|01\right\rangle+c_1\left|10\right\rangle-d_1\left|11\right\rangle$&&&$I\otimes\sigma_z$ \\[1ex]
$\left|\Phi^+\right\rangle\left|\Phi^+\right\rangle$&&&$\left|00\right\rangle-\left|11\right\rangle$&&&$	a_1\left|00\right\rangle+b_1\left|01\right\rangle+c_1\left|10\right\rangle+d_1\left|11\right\rangle$&&&$I\otimes I$ \\[1ex]
$\left|\Phi^-\right\rangle\left|\Phi^+\right\rangle$&&&$\left|00\right\rangle-\left|11\right\rangle$&&&$	a_1\left|00\right\rangle+b_1\left|01\right\rangle-c_1\left|10\right\rangle-d_1\left|11\right\rangle$&&&$\sigma_z\otimes I$ \\[1ex]
$\left|\Phi^+\right\rangle\left|\Phi^-\right\rangle$&&&$\left|00\right\rangle-\left|11\right\rangle$&&&$	a_1\left|00\right\rangle-b_1\left|01\right\rangle+c_1\left|10\right\rangle-d_1\left|11\right\rangle$&&&$I\otimes\sigma_z$ \\[1ex]
$\left|\Phi^-\right\rangle\left|\Phi^-\right\rangle$&&&$\left|00\right\rangle-\left|11\right\rangle$&&&$	a_1\left|00\right\rangle-b_1\left|01\right\rangle-c_1\left|10\right\rangle+d_1\left|11\right\rangle$&&&$\sigma_z\otimes\sigma_z$ \\[0.5ex]\hline\\[0.2ex]
$\left|\Phi^+\right\rangle\left|\Phi^+\right\rangle$&&&$\left|01\right\rangle+\left|10\right\rangle$&&&$	a_1\left|10\right\rangle+b_1\left|11\right\rangle+c_1\left|10\right\rangle+d_1\left|01\right\rangle$&&&$\sigma_x\otimes I$ \\[1ex]
$\left|\Phi^-\right\rangle\left|\Phi^+\right\rangle$&&&$\left|01\right\rangle+\left|10\right\rangle$&&&$	a_1\left|10\right\rangle+b_1\left|11\right\rangle-c_1\left|10\right\rangle-d_1\left|01\right\rangle$&&&$-i\sigma_y\otimes I$ \\[1ex]
$\left|\Phi^+\right\rangle\left|\Phi^-\right\rangle$&&&$\left|01\right\rangle+\left|10\right\rangle$&&&$	a_1\left|10\right\rangle-b_1\left|11\right\rangle+c_1\left|10\right\rangle-d_1\left|01\right\rangle$&&&$\sigma_x\otimes\sigma_z$ \\[1ex]
$\left|\Phi^-\right\rangle\left|\Phi^-\right\rangle$&&&$\left|01\right\rangle+\left|10\right\rangle$&&&$	a_1\left|10\right\rangle-b_1\left|11\right\rangle-c_1\left|10\right\rangle+d_1\left|01\right\rangle$&&&$-i\sigma_y\otimes\sigma_z$ \\[1ex]
$\left|\Phi^+\right\rangle\left|\Phi^+\right\rangle$&&&$\left|01\right\rangle-\left|10\right\rangle$&&&$	-a_1\left|10\right\rangle-b_1\left|11\right\rangle+c_1\left|10\right\rangle+d_1\left|01\right\rangle$&&&$i\sigma_y\otimes I$ \\[1ex]
$\left|\Phi^-\right\rangle\left|\Phi^+\right\rangle$&&&$\left|01\right\rangle-\left|10\right\rangle$&&&$	-a_1\left|10\right\rangle-b_1\left|11\right\rangle-c_1\left|10\right\rangle-d_1\left|01\right\rangle$&&&$-\sigma_x\otimes I$ \\[1ex]
$\left|\Phi^+\right\rangle\left|\Phi^-\right\rangle$&&&$\left|01\right\rangle-\left|10\right\rangle$&&&$	-a_1\left|10\right\rangle+b_1\left|11\right\rangle+c_1\left|10\right\rangle-d_1\left|01\right\rangle$&&&$i\sigma_y\otimes\sigma_z$ \\[1ex]
$\left|\Phi^-\right\rangle\left|\Phi^-\right\rangle$&&&$\left|01\right\rangle-\left|10\right\rangle$&&&$	-a_1\left|10\right\rangle+b_1\left|11\right\rangle-c_1\left|10\right\rangle+d_1\left|01\right\rangle$&&&$-\sigma_x\otimes\sigma_z$ \\[0.5ex]\hline\\[0.5ex]
$\left|\Phi^+\right\rangle\left|\Psi^+\right\rangle$&&&$\left|00\right\rangle+\left|11\right\rangle$&&&$	a_1\left|01\right\rangle+b_1\left|00\right\rangle-c_1\left|11\right\rangle-d_1\left|10\right\rangle$&&&$\sigma_z\otimes\sigma_x$ \\[1ex]
$\left|\Phi^-\right\rangle\left|\Psi^+\right\rangle$&&&$\left|00\right\rangle+\left|11\right\rangle$&&&$	a_1\left|01\right\rangle+b_1\left|00\right\rangle+c_1\left|11\right\rangle+d_1\left|10\right\rangle$&&&$I\otimes\sigma_x$ \\[1ex]
$\left|\Phi^+\right\rangle\left|\Psi^-\right\rangle$&&&$\left|00\right\rangle+\left|11\right\rangle$&&&$	a_1\left|01\right\rangle-b_1\left|00\right\rangle-c_1\left|11\right\rangle+d_1\left|10\right\rangle$&&&$\sigma_z\otimes-i\sigma_y$ \\[1ex]
$\left|\Phi^-\right\rangle\left|\Psi^-\right\rangle$&&&$\left|00\right\rangle+\left|11\right\rangle$&&&$	a_1\left|01\right\rangle-b_1\left|00\right\rangle+c_1\left|11\right\rangle-d_1\left|10\right\rangle$&&&$I\otimes-i\sigma_y$ \\[1ex]
$\left|\Phi^+\right\rangle\left|\Psi^+\right\rangle$&&&$\left|00\right\rangle-\left|11\right\rangle$&&&$	a_1\left|01\right\rangle+b_1\left|00\right\rangle+c_1\left|11\right\rangle+d_1\left|10\right\rangle$&&&$I\otimes\sigma_x$ \\[1ex]
$\left|\Phi^-\right\rangle\left|\Psi^+\right\rangle$&&&$\left|00\right\rangle-\left|11\right\rangle$&&&$	a_1\left|01\right\rangle+b_1\left|00\right\rangle-c_1\left|11\right\rangle-d_1\left|10\right\rangle$&&&$\sigma_z\otimes\sigma_x$ \\[1ex]
$\left|\Phi^+\right\rangle\left|\Psi^-\right\rangle$&&&$\left|00\right\rangle-\left|11\right\rangle$&&&$	a_1\left|01\right\rangle-b_1\left|00\right\rangle+c_1\left|11\right\rangle-d_1\left|10\right\rangle$&&&$I\otimes-i\sigma_y$ \\[1ex]
$\left|\Phi^-\right\rangle\left|\Psi^-\right\rangle$&&&$\left|00\right\rangle-\left|11\right\rangle$&&&$	a_1\left|01\right\rangle-b_1\left|00\right\rangle-c_1\left|11\right\rangle+d_1\left|10\right\rangle$&&&$\sigma_z\otimes-i\sigma_y$ \\[0.5ex]\hline\\[0.5ex]
$\left|\Phi^+\right\rangle\left|\Psi^+\right\rangle$&&&$\left|01\right\rangle+\left|10\right\rangle$&&&$	a_1\left|11\right\rangle+b_1\left|10\right\rangle+c_1\left|01\right\rangle+d_1\left|00\right\rangle$&&&$\sigma_x\otimes\sigma_x$ \\[1ex]
$\left|\Phi^-\right\rangle\left|\Psi^+\right\rangle$&&&$\left|01\right\rangle+\left|10\right\rangle$&&&$	a_1\left|11\right\rangle+b_1\left|10\right\rangle-c_1\left|01\right\rangle-d_1\left|00\right\rangle$&&&$-i\sigma_y\otimes\sigma_x$ \\[1ex]
$\left|\Phi^+\right\rangle\left|\Psi^-\right\rangle$&&&$\left|01\right\rangle+\left|10\right\rangle$&&&$	a_1\left|11\right\rangle-b_1\left|10\right\rangle+c_1\left|01\right\rangle-d_1\left|00\right\rangle$&&&$\sigma_x\otimes-i\sigma_y$ \\[1ex]
$\left|\Phi^-\right\rangle\left|\Psi^-\right\rangle$&&&$\left|01\right\rangle+\left|10\right\rangle$&&&$	a_1\left|11\right\rangle-b_1\left|10\right\rangle-c_1\left|01\right\rangle+d_1\left|00\right\rangle$&&&$-i\sigma_y\otimes-i\sigma_y$ \\[1ex]
$\left|\Phi^+\right\rangle\left|\Psi^+\right\rangle$&&&$\left|01\right\rangle-\left|10\right\rangle$&&&$	-a_1\left|11\right\rangle-b_1\left|10\right\rangle+c_1\left|01\right\rangle+d_1\left|00\right\rangle$&&&$i\sigma_y\otimes\sigma_x$ \\[1ex]
$\left|\Phi^-\right\rangle\left|\Psi^+\right\rangle$&&&$\left|01\right\rangle-\left|10\right\rangle$&&&$	-a_1\left|11\right\rangle-b_1\left|10\right\rangle-c_1\left|01\right\rangle-d_1\left|00\right\rangle$&&&$-\sigma_x\otimes\sigma_x$ \\[1ex]
$\left|\Phi^+\right\rangle\left|\Psi^-\right\rangle$&&&$\left|01\right\rangle-\left|10\right\rangle$&&&$	-a_1\left|11\right\rangle+b_1\left|10\right\rangle+c_1\left|01\right\rangle-d_1\left|00\right\rangle$&&&$-i\sigma_y\otimes i\sigma_y$ \\[1ex]
$\left|\Phi^-\right\rangle\left|\Psi^-\right\rangle$&&&$\left|01\right\rangle-\left|10\right\rangle$&&&$	-a_1\left|11\right\rangle+b_1\left|10\right\rangle-c_1\left|01\right\rangle+d_1\left|00\right\rangle$&&&$\sigma_x\otimes i\sigma_y$ \\[1ex]\hline\hline
\end{tabular}\label{tab1}}}
\end{table}

\begin{table}[!h]
\setcounter{table}{0}
\renewcommand{\thetable}{1 continued}
{
\caption{\footnotesize} \vspace*{10pt}{\footnotesize
\begin{tabular}{cccccccccc}\hline\hline
{}&{}&{}&{}{}&{}&{}&{}&{}&{}\\[-1.0ex]
Alice's result&&&Bob's result&&& State obtained by Chika&&&Unitary transformation\\[1ex]\hline
$\left|\Psi^+\right\rangle\left|\Phi^+\right\rangle$&&&$\left|00\right\rangle+\left|11\right\rangle$&&&$	-a_1\left|10\right\rangle-b_1\left|11\right\rangle+c_1\left|00\right\rangle+d_1\left|01\right\rangle$&&&$i\sigma_y\otimes I$ \\[1ex]
$\left|\Psi^-\right\rangle\left|\Phi^+\right\rangle$&&&$\left|00\right\rangle+\left|11\right\rangle$&&&$	-a_1\left|10\right\rangle-b_1\left|11\right\rangle-c_1\left|00\right\rangle-d_1\left|01\right\rangle$&&&$-\sigma_x\otimes I$ \\[1ex]
$\left|\Psi^+\right\rangle\left|\Phi^-\right\rangle$&&&$\left|00\right\rangle+\left|11\right\rangle$&&&$	-a_1\left|10\right\rangle+b_1\left|11\right\rangle+c_1\left|00\right\rangle-d_1\left|01\right\rangle$&&&$i\sigma_y\otimes\sigma_z$ \\[1ex]
$\left|\Psi^-\right\rangle\left|\Phi^-\right\rangle$&&&$\left|00\right\rangle+\left|11\right\rangle$&&&$	-a_1\left|10\right\rangle+b_1\left|11\right\rangle-c_1\left|00\right\rangle+d_1\left|01\right\rangle$&&&$-\sigma_x\otimes\sigma_z$ \\[1ex]
$\left|\Psi^+\right\rangle\left|\Phi^+\right\rangle$&&&$\left|00\right\rangle-\left|11\right\rangle$&&&$	a_1\left|10\right\rangle+b_1\left|11\right\rangle+c_1\left|00\right\rangle+d_1\left|01\right\rangle$&&&$\sigma_x\otimes I$ \\[1ex]
$\left|\Psi^-\right\rangle\left|\Phi^+\right\rangle$&&&$\left|00\right\rangle-\left|11\right\rangle$&&&$	a_1\left|10\right\rangle+b_1\left|11\right\rangle-c_1\left|00\right\rangle-d_1\left|01\right\rangle$&&&$-i\sigma_y\otimes I$ \\[1ex]
$\left|\Psi^+\right\rangle\left|\Phi^-\right\rangle$&&&$\left|00\right\rangle-\left|11\right\rangle$&&&$	a_1\left|10\right\rangle-b_1\left|11\right\rangle+c_1\left|00\right\rangle-d_1\left|01\right\rangle$&&&$\sigma_x\otimes\sigma_z$ \\[1ex]
$\left|\Psi^-\right\rangle\left|\Phi^-\right\rangle$&&&$\left|00\right\rangle-\left|11\right\rangle$&&&$	a_1\left|10\right\rangle-b_1\left|11\right\rangle-c_1\left|00\right\rangle+d_1\left|01\right\rangle$&&&$-i\sigma_y\otimes\sigma_z$ \\[0.5ex]\hline\\[0.5ex]
$\left|\Psi^+\right\rangle\left|\Phi^+\right\rangle$&&&$\left|01\right\rangle+\left|11\right\rangle$&&&$	a_1\left|00\right\rangle+b_1\left|01\right\rangle+c_1\left|10\right\rangle+d_1\left|11\right\rangle$&&&$I\otimes I$ \\[1ex]
$\left|\Psi^-\right\rangle\left|\Phi^+\right\rangle$&&&$\left|01\right\rangle+\left|11\right\rangle$&&&$	a_1\left|00\right\rangle+b_1\left|01\right\rangle-c_1\left|10\right\rangle-d_1\left|11\right\rangle$&&&$\sigma_z\otimes I$ \\[1ex]
$\left|\Psi^+\right\rangle\left|\Phi^-\right\rangle$&&&$\left|01\right\rangle+\left|11\right\rangle$&&&$	a_1\left|00\right\rangle-b_1\left|01\right\rangle+c_1\left|10\right\rangle-d_1\left|11\right\rangle$&&&$I\otimes\sigma_z$ \\[1ex]
$\left|\Psi^-\right\rangle\left|\Phi^-\right\rangle$&&&$\left|01\right\rangle+\left|11\right\rangle$&&&$	a_1\left|00\right\rangle-b_1\left|01\right\rangle-c_1\left|10\right\rangle+d_1\left|11\right\rangle$&&&$\sigma_z\otimes\sigma_z$ \\[1ex]
$\left|\Psi^+\right\rangle\left|\Phi^+\right\rangle$&&&$\left|01\right\rangle-\left|11\right\rangle$&&&$	a_1\left|00\right\rangle+b_1\left|01\right\rangle-c_1\left|10\right\rangle-d_1\left|11\right\rangle$&&&$\sigma_z\otimes I$ \\[1ex]
$\left|\Psi^-\right\rangle\left|\Phi^+\right\rangle$&&&$\left|01\right\rangle-\left|11\right\rangle$&&&$	a_1\left|00\right\rangle+b_1\left|01\right\rangle+c_1\left|10\right\rangle+d_1\left|11\right\rangle$&&&$I\otimes I$ \\[1ex]
$\left|\Psi^+\right\rangle\left|\Phi^-\right\rangle$&&&$\left|01\right\rangle-\left|11\right\rangle$&&&$	a_1\left|00\right\rangle-b_1\left|01\right\rangle-c_1\left|10\right\rangle+d_1\left|11\right\rangle$&&&$\sigma_z\otimes\sigma_z$ \\[1ex]
$\left|\Psi^-\right\rangle\left|\Phi^-\right\rangle$&&&$\left|01\right\rangle-\left|11\right\rangle$&&&$	a_1\left|00\right\rangle-b_1\left|01\right\rangle+c_1\left|10\right\rangle-d_1\left|11\right\rangle$&&&$I\otimes\sigma_z$ \\[0.5ex]\hline\\[0.5ex]
$\left|\Psi^+\right\rangle\left|\Psi^+\right\rangle$&&&$\left|00\right\rangle+\left|11\right\rangle$&&&$	-a_1\left|11\right\rangle-b_1\left|10\right\rangle+c_1\left|01\right\rangle+d_1\left|00\right\rangle$&&&$i\sigma_y\otimes\sigma_x$ \\[1ex]
$\left|\Psi^-\right\rangle\left|\Psi^+\right\rangle$&&&$\left|00\right\rangle+\left|11\right\rangle$&&&$	-a_1\left|11\right\rangle-b_1\left|10\right\rangle-c_1\left|01\right\rangle-d_1\left|00\right\rangle$&&&$-\sigma_x\otimes\sigma_x$ \\[1ex]
$\left|\Psi^+\right\rangle\left|\Psi^-\right\rangle$&&&$\left|00\right\rangle+\left|11\right\rangle$&&&$	-a_1\left|11\right\rangle+b_1\left|10\right\rangle+c_1\left|01\right\rangle-d_1\left|00\right\rangle$&&&$-i\sigma_y\otimes i\sigma_y$ \\[1ex]
$\left|\Psi^-\right\rangle\left|\Psi^-\right\rangle$&&&$\left|00\right\rangle+\left|11\right\rangle$&&&$	-a_1\left|11\right\rangle+b_1\left|10\right\rangle-c_1\left|01\right\rangle+d_1\left|00\right\rangle$&&&$\sigma_x\otimes i\sigma_y$ \\[1ex]
$\left|\Psi^+\right\rangle\left|\Psi^+\right\rangle$&&&$\left|00\right\rangle-\left|11\right\rangle$&&&$	a_1\left|11\right\rangle+b_1\left|10\right\rangle+c_1\left|01\right\rangle+d_1\left|00\right\rangle$&&&$\sigma_x\otimes\sigma_x$ \\[1ex]
$\left|\Psi^-\right\rangle\left|\Psi^+\right\rangle$&&&$\left|00\right\rangle-\left|11\right\rangle$&&&$	a_1\left|11\right\rangle+b_1\left|10\right\rangle-c_1\left|01\right\rangle-d_1\left|00\right\rangle$&&&$-i\sigma_y\otimes\sigma_x$ \\[1ex]
$\left|\Psi^+\right\rangle\left|\Psi^-\right\rangle$&&&$\left|00\right\rangle-\left|11\right\rangle$&&&$	a_1\left|11\right\rangle-b_1\left|10\right\rangle+c_1\left|01\right\rangle-d_1\left|00\right\rangle$&&&$\sigma_x\otimes-i\sigma_y$ \\[1ex]
$\left|\Psi^-\right\rangle\left|\Psi^-\right\rangle$&&&$\left|00\right\rangle-\left|11\right\rangle$&&&$	a_1\left|11\right\rangle-b_1\left|10\right\rangle-c_1\left|01\right\rangle+d_1\left|00\right\rangle$&&&$-i\sigma_y\otimes-i\sigma_y$ \\[0.5ex]\hline\\[0.5ex]
$\left|\Psi^+\right\rangle\left|\Psi^+\right\rangle$&&&$\left|01\right\rangle+\left|10\right\rangle$&&&$	a_1\left|01\right\rangle+b_1\left|00\right\rangle+c_1\left|11\right\rangle+d_1\left|10\right\rangle$&&&$I\otimes\sigma_x$ \\[1ex]
$\left|\Psi^-\right\rangle\left|\Psi^+\right\rangle$&&&$\left|01\right\rangle+\left|10\right\rangle$&&&$	a_1\left|01\right\rangle+b_1\left|00\right\rangle-c_1\left|11\right\rangle-d_1\left|10\right\rangle$&&&$\sigma_z\otimes\sigma_x$ \\[1ex]
$\left|\Psi^+\right\rangle\left|\Psi^-\right\rangle$&&&$\left|01\right\rangle+\left|10\right\rangle$&&&$	a_1\left|01\right\rangle-b_1\left|00\right\rangle+c_1\left|11\right\rangle-d_1\left|10\right\rangle$&&&$I\otimes-i\sigma_y$ \\[1ex]
$\left|\Psi^-\right\rangle\left|\Psi^-\right\rangle$&&&$\left|01\right\rangle+\left|10\right\rangle$&&&$	a_1\left|01\right\rangle-b_1\left|00\right\rangle-c_1\left|11\right\rangle+d_1\left|10\right\rangle$&&&$\sigma_z\otimes-i\sigma_y$ \\[1ex]
$\left|\Psi^+\right\rangle\left|\Psi^+\right\rangle$&&&$\left|01\right\rangle-\left|10\right\rangle$&&&$	a_1\left|01\right\rangle+b_1\left|00\right\rangle-c_1\left|11\right\rangle-d_1\left|10\right\rangle$&&&$\sigma_z\otimes\sigma_x$ \\[1ex]
$\left|\Psi^-\right\rangle\left|\Psi^+\right\rangle$&&&$\left|01\right\rangle-\left|10\right\rangle$&&&$	a_1\left|01\right\rangle+b_1\left|00\right\rangle+c_1\left|11\right\rangle+d_1\left|10\right\rangle$&&&$I\otimes\sigma_x$ \\[1ex]
$\left|\Psi^+\right\rangle\left|\Psi^-\right\rangle$&&&$\left|01\right\rangle-\left|10\right\rangle$&&&$	a_1\left|01\right\rangle-b_1\left|00\right\rangle-c_1\left|11\right\rangle+d_1\left|10\right\rangle$&&&$\sigma_z\otimes-i\sigma_y$ \\[1ex]
$\left|\Psi^-\right\rangle\left|\Psi^-\right\rangle$&&&$\left|01\right\rangle-\left|10\right\rangle$&&&$	a_1\left|01\right\rangle-b_1\left|00\right\rangle+c_1\left|11\right\rangle-d_1\left|10\right\rangle$&&&$I\otimes-i\sigma_y$ \\[1ex]
\hline\hline\end{tabular}\label{tab2}}
\vspace*{-1pt}}
\end{table}
\section{Conclusion}
In summary, this paper demonstrates a tripartite scheme for the probabilistic teleportation of an arbitrary single qubit state, without losing the information of the state being teleported, via a four-qubit cluster state \cite{B1} as the quantum channel linking the three parties. The sender performs BSM and communicates the results of the measurement to the controller, who also performs BSM on his qubit pairs. With the introduction of an auxiliary qubit state, using a suitable unitary transformation and a POVM, the receiver can recreate the state of the original qubit. We also studied quantum information splitting of an arbitrary two-particle system via a four-qubit cluster state and a Bell state as the quantum channel. Problems related to security attacks were examined, and as we have shown, this protocol is robust to any attack. This demonstration is another supportive evidence that justifies quantum entanglement as the key resource in quantum information science, and it also represents the continuation of Refs. \cite{R3,T33}.
\newpage
\section*{Acknowledgments}
We thank the referees for the positive enlightening comments and suggestions, which have greatly helped us in making improvements to this paper. In addition, B.J.F. acknowledges Prof. K. J. Oyewumi for his unending support.

\end{document}